\documentclass[preprint,12pt]{elsarticle}

\usepackage{amssymb}

\begin{document}

\begin{frontmatter}



\title{Anderson localization and saturable nonlinearity in one-dimensional disordered lattices}

\author[bpn1,bpn2]{Ba Phi Nguyen}
\ead{nguyenbaphi@muce.edu.vn}
\address[bpn1]{Department of Basic Sciences, Mientrung University of Civil Engineering, 24 Nguyen Du, Tuy Hoa, Vietnam}
\address[bpn2]{Institute of NT-IT Fusion Technology, Ajou University, Suwon 16499, Korea}
\author[kk]{Kihong Kim\corref{cor2}}
\cortext[cor2]{Corresponding author}
\ead{khkim@ajou.ac.kr}
\address[kk]{Department of Energy Systems Research and Department of Physics, Ajou University, Suwon 16499, Korea}

\begin{abstract}
We investigate numerically the propagation and the Anderson localization of plane waves in a one-dimensional lattice chain, where disorder and saturable nonlinearity are simultaneously present. Using a calculation scheme for solving the stationary discrete nonlinear Schr\"{o}dinger equation in the fixed input case, the disorder-averaged logarithmic transmittance and the localization length are calculated in a numerically precise manner. The localization length is found to be a nonmonotonic function of the incident wave intensity, acquiring a minimum value at a certain finite intensity, due to saturation effects. For low incident intensities where the saturation effect is ineffective, the enhancement of localization due to Kerr-type nonlinearity occurs in a way similar to the case without saturation. For sufficiently high incident intensities, we find that the localization length is an increasing function of the incident wave intensity, which implies that localization is suppressed for stronger input intensities, and ultimately approaches a saturation value. This feature is associated with the fact that the nonlinear system is reduced to an effectively linear one, when either the incident wave intensity or the saturation parameter is sufficiently large.
We also find that the nonlinear saturation effect is stronger and more pronounced when the energy of the incident wave is larger.
\end{abstract}

\begin{keyword}
Anderson localization \sep Saturable nonlinearity \sep Localization length



\end{keyword}

\end{frontmatter}


\section{Introduction}

Understanding the influence of nonlinearity on Anderson localization of waves in disordered systems has attracted much interest from researchers over the past years \cite{Dev,Do1,Do2,Gre,Mol,Ras,Hen,Tro,Ng1,Ng2,Ko11,Ko12,Ko13,Shep,Kot,Ko2,Pik,Lah,Fla}. In nonlinear wave propagation problems, all aspects of the dynamics are not determined by the spectral properties unlike in linear problems. Due to the peculiarity of nonlinear systems, different ways of posing the problem are not equivalent to each other \cite{McK}. As a consequence, there exist many uncertainties and unanswered questions associated with the physical systems in which disorder and nonlinearity are simultaneously present \cite{Fis,Maf}. Within this context, a fundamental question is whether the presence of nonlinearity enhances or suppresses Anderson localization. The answer to this question can best be described as inconclusive at this point. The influence of nonlinearity on Anderson localization can be studied in three different contexts: (i) the transmission of plane waves through disordered nonlinear media \cite{Dev,Do1,Do2,Gre,Mol,Ras,Hen,Tro,Ng1,Ng2}, (ii) the effect of nonlinear perturbations on localized eigenstates in finite-size systems \cite{Ko11,Ko12,Ko13,Lah}, (iii) the effect of nonlinearity on the spreading of initially localized wave packets \cite{Mol,Shep,Kot,Ko2,Pik,Fla}. The discrete nonlinear Schr\"{o}dinger equation can be used in the study of all these cases \cite{Kev}.

Traditionally, much interest has been given to the standard cubic nonlinearity associated with the Kerr effect due to its relevance to a wide range of physical systems \cite{Dev,Do1,Do2,Gre,Mol,Ras,Hen,Tro,Ng1,Ng2,Ko11,Ko12,Ko13,Shep,Kot,Ko2,Pik,Lah,Fla}. In particular, it has been shown that the effect of nonlinearity on Anderson localization is qualitatively different for localized and extended excitations. It favors the propagation of localized excitations, while it inhibits that of extended excitations \cite{Mol}. By studying localized eigenstates and wave packet expansion in one-dimensional (1D) disordered lattices, Lahini {\it et al.} have observed that in the weakly nonlinear regime, nonlinearity enhances localization for flat-phased states and induces delocalization for staggered states \cite{Lah}. Other papers have also reported that pure Anderson localization is destroyed and turns into a subdiffusive spreading of wave packets in the presence of nonlinearity \cite{Ko2,Pik,Fla}.

However, the cubic nonlinearity does not always reflect the physical reality and, in certain cases, other kinds of nonlinearity should be considered. For example, for short and high peak power pulses, the field-induced change in the refractive index cannot be described by a Kerr-type nonlinearity, since it is influenced by higher-order nonlinearities such as saturable nonlinearity \cite{Gat}. The saturation of the nonlinear response has been shown to have novel impacts on the dynamical properties of wave propagation in clean systems \cite{Gat,Cue,Mel,Vin,Led,Nae,Guz,Sam,Hu,Cao,Ass,Law}. For instance, it has been pointed out that the saturable nonlinearity allows the existence of stable two-state solitons with the same time duration \cite{Gat} and of breathers with high power \cite{Cue}. In \cite{Mel}, genuinely localized travelling waves have been found in saturable nonlinear Schr\"{o}dinger lattices for the first time. Furthermore, in recent papers, the authors have shown that there exists asymmetric wave propagation through saturable nonlinear oligomers and wave rectification devices can be produced based on such systems \cite{Ass,Law}.

Until now, many researches have been devoted to soliton propagation in saturable nonlinear systems. To the best of our knowledge, there has been very little research on the interplay between saturable nonlinearity and disorder \cite{Li,Rad,Dos}. The authors of \cite{Li} have investigated light localization in disordered photo-refractive lattices with saturable self-defocusing nonlinearity. It has been demonstrated that by increasing the value of the nonlinearity strength continuously, a phase transition from a localized state to an extended state can be obtained. In a very recent paper, the authors have studied numerically the dynamics of an initially localized wave packet in 1D disordered chains with saturable nonlinearity \cite{Dos}. From a detailed numerical analysis, they have found that saturable nonlinearity can promote a sub-diffusive spreading of the wave packet even in the presence of diagonal disorder for a long time. In addition, they have also investigated the effect of saturated nonlinearity for initial times of the electronic evolution thus showing the possibility of mobile breather-like modes.

In this work, we investigate the transmission and localization properties of plane waves in a 1D disordered lattice chain with saturable nonlinearity. Using a disordered version of the stationary discrete nonlinear Schr\"{o}dinger equation, the disorder-averaged logarithmic transmittance and the localization length are calculated in a numerically precise manner. Unlike in many previous works, we strictly fix the intensity of the incident wave and compute physical quantities as a function of other parameters. We find that the localization length is a nonmonotonic function of the incident wave intensity. For low incident intensities, the enhancement of localization due to nonlinearity occurs in a way similar to the case without saturation. For sufficiently high incident intensities, we find that the localization length is an increasing function of the incident wave intensity and ultimately approaches a saturation value.
We also find that the nonlinear saturation effect is stronger and more pronounced when the energy of the incident wave is larger.

\section{Theoretical model and numerical method}
\subsection{Theoretical model}

In order to study the effects of saturable nonlinearity on Anderson localization in a 1D disordered lattice, we use the discrete nonlinear Schr\"{o}dinger equation given by
\begin{eqnarray}
i\hbar\frac{dC_{n}(t)}{dt}=\epsilon_{n}C_{n}(t)-V\left[C_{n+1}(t)+C_{n-1}(t)\right]\nonumber\\
+\frac{\alpha\vert C_{n}(t)\vert ^2}{1+\beta\vert C_{n}(t)\vert^2}C_{n}(t),
\label{equation1}
\end{eqnarray}
where $C_{n}(t)$ is the wave function amplitude and $V$ is the hopping integral between the nearest-neighbor sites. The on-site potential $\epsilon_{n}$ varies randomly as a function of the site index $n$. We assume that $\epsilon_n$'s are independent random variables distributed uniformly in the range $[-W/2, W/2]$. The parameter $\alpha$ is the strength of a third-order Kerr-type nonlinear response in the regime of low-intensity waves, while $\beta$ is the degree of saturation of the nonlinearity. We set $\hbar=V=1$ for convenience. We note that when the intensity of the incident wave is sufficiently weak so that $\vert C_{n}(t)\vert ^2\ll 1$, the last term of Eq.~(\ref{equation1}) can be approximated by the expression $[\alpha\vert C_{n}(t)\vert^2C_{n}(t)-\alpha\beta\vert C_{n}(t)\vert ^4C_{n}(t)]$. With this approximation, we recover the standard cubic discrete nonlinear Schr\"{o}dinger equation with an additional quartic term $-\alpha\beta\vert C_{n}(t)\vert ^4C_{n}(t)$.

In a recent paper, the authors have studied the influence of saturable nonlinearity on Anderson localization of an initially localized wave packet
by integrating Eq.~(\ref{equation1}) directly \cite{Dos}. In the present work, we focus on the transmission problem and study how an incident plane wave is transmitted through a finite segment of a saturable nonlinear disordered medium. Such a setup may be realized in experiments on layered photonic structures, where wave transmission through a nonlinear disordered medium can be studied as a function of the intensity of the incident wave. To this end, we will look for stationary solutions to Eq.~(\ref{equation1}) of the form $C_{n}(t)=\psi_{n}e^{-iEt}$, where $E$ represents the energy of the incoming wave. This leads to a set of coupled algebraic equations for the complex amplitude $\psi_{n}$:
\begin{eqnarray}
E\psi_{n}=\epsilon_{n}\psi_{n}-\psi_{n+1}-\psi_{n-1}+\frac{\alpha\vert\psi_{n}\vert^2}{1+\beta\vert\psi_{n}\vert^2}\psi_{n}.
\label{equation2}
\end{eqnarray}

In order to allow the free propagation of the wave outside the disordered nonlinear region, we take $\epsilon_{n}$, $\alpha$ and $\beta$ to be nonzero only in the region $1 \leq n\leq L$. We note that the transformation $C_{n}\rightarrow (-1)^n C^{\ast}_{n}$, $\alpha\rightarrow -\alpha$ and $\epsilon_{n}\rightarrow -\epsilon_{n}$ leaves Eq.~(\ref{equation1}) invariant \cite{Tie}. Therefore the sign of $\alpha$ can be fixed to be positive when we consider disorder-averaged quantities.

\begin{figure}
\resizebox{0.7\textwidth}{!}{%
\includegraphics{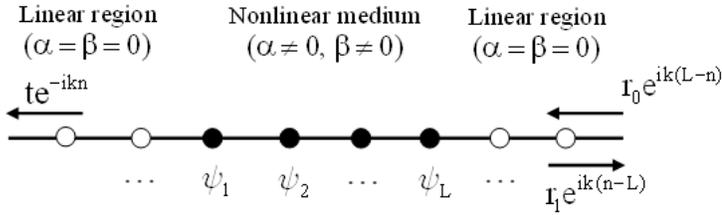}}
\caption{Schematic of the transmission problem. A plane wave is incident on a saturable nonlinear disordered medium with $\alpha \neq 0$ and $\beta \neq 0$ from a uniform linear region on the right ($n>L$) and is transmitted to a uniform linear region on the left ($n<1$). $r_{0}$, $r_{1}$ and $t$ are the complex amplitudes of the incident, reflected and transmitted waves respectively.}
\end{figure}

Let us assume that a plane wave is incident from the right side (see Fig.~1) and define the complex amplitudes of the incident, reflected and transmitted waves, $r_{0}$, $r_{1}$ and $t$, by
\begin{eqnarray}
\psi_n=\left\{\begin{array}{l l}
r_{0}e^{ik(L-n)}+r_{1}e^{ik(n-L)}, & \quad \mbox{$n\geq L$},\\
te^{-ikn}, & \quad \mbox{$n\leq 0$,}
\end{array}\right.
\label{equation3}
\end{eqnarray}
where the wave number $k$ is related to $E$ by $E=-2\cos k$. In the absence of dissipation, the
conservation of probability current requires that $\vert
r_{1}\vert^2+ \vert t\vert^2= \vert r_{0}\vert^2$. We choose the
overall constant phase for the wave functions so that $t$ is a
positive real number.

\subsection{Numerical method: fixed input problem}

In order to solve Eqs.~(\ref{equation2}) and (\ref{equation3})
numerically in the situation where the intensity of the input wave is fixed,
we first choose a certain positive real number for $t$.
Then we use the relationships $\psi_{-1}=te^{ik}$, $\psi_0=t$ and
$\psi_1=-Et-te^{ik}=te^{-ik}$, which follows from Eqs.~(\ref{equation2}) and (\ref{equation3}), and solve Eq.~(\ref{equation2}) iteratively
until we obtain $\psi_L$ and $\psi_{L+1}$ for a given $L$. Next,
using the relationships $\psi_L=r_0+r_1$ and
$\psi_{L+1}=r_0e^{-ik}+r_1 e^{ik}$, we write $r_0$ and $r_1$ as
\begin{eqnarray}
r_0=\frac{\psi_L e^{ik}-\psi_{L+1}}{e^{ik}-e^{-ik}},~~
r_1=\frac{\psi_{L+1}-\psi_{L}e^{-ik}}{e^{ik}-e^{-ik}}.
\label{equation4}
\end{eqnarray}
The reflectance $R$ and the transmittance $T$
are expressed as
\begin{eqnarray}
R&=&\bigg\vert\frac{r_{1}}{r_{0}}\bigg\vert^2= \frac{\vert
\psi_{L}e^{-ik}-\psi_{L+1}\vert ^2}{\vert \psi_L
e^{ik}-\psi_{L+1}\vert
^2},\nonumber\\
T&=&\bigg\vert\frac{t}{r_{0}}\bigg\vert^2=\vert t\vert^2 \frac{4
{\sin}^2k}{\vert \psi_L e^{ik}-\psi_{L+1}\vert^2}.
\label{equation5}
\end{eqnarray}

In the linear case, the values of $R$ and $T$ are independent of the
initial choice of $t$. This property, however, does not hold in the
nonlinear problem. The case where the value of
$\vert t\vert^2$ is fixed is called the fixed output case, while
that with a fixed value of $\vert r_0\vert^2$ is called the fixed
input case. These two problems are fundamentally different and need to
be carefully distinguished. In the fixed output
case, the nonlinear wave equation is known to have a unique solution, whereas in
the fixed input case, it can have multiple solutions and the
phenomenon of optical bistability or multistability can occur. In
actual experiments, one usually fixes the strength of the input wave
while varying other parameters. In the present work, we consider
only the fixed input case.

In order to calculate $R$ and $T$ in the
fixed input problem, we first choose the values of $L$, $E$ (or
$k$), $\alpha$, $\beta$, $W$ and $\vert r_0\vert^2$ and a random
configuration of $\epsilon_n$. Then we solve Eq.~(\ref{equation2})
repeatedly for many different initial values of $t$ ($t = 0$,
$\delta$, $2\delta$, $3\delta$, $\cdots$) until we obtain the final
value of $\vert r_0\vert^2$ sufficiently close to the pre-chosen value.
The step size $\delta$ needs to be chosen appropriately to achieve
desired accuracy. In general, there may be several values of $t$
which gives the same value of $\vert r_0\vert^2$. In the present
study, we stick to only the first solution corresponding to the
smallest value of $t$.

In the study of the Anderson localization phenomenon of waves in
disordered media, the central quantity to calculate is the localization
length $\xi$ defined by
\begin{eqnarray}
\frac{1}{\xi}=-\lim_{L\to \infty}\frac{\langle \ln T \rangle}{L},
\label{equation6}
\end{eqnarray}
where $\langle \cdots \rangle$ denotes averaging over a large number of random site energy realizations.

\section{Results}

\begin{figure}
\resizebox{0.7\textwidth}{!}{%
\includegraphics{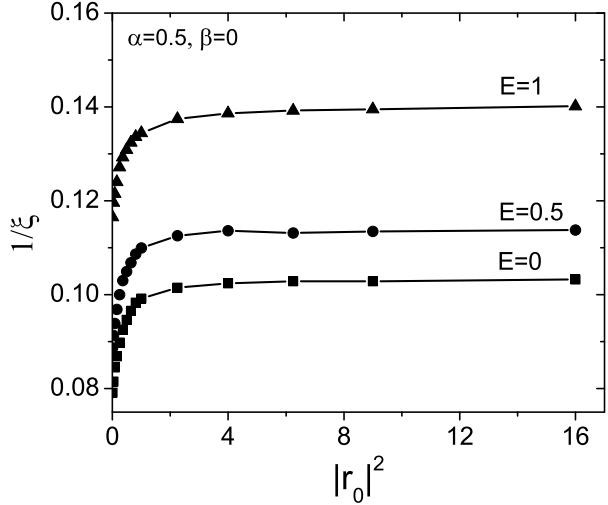}}
\caption{Inverse localization length $1/\xi$ plotted as a function of the incident wave intensity $\vert r_0\vert^2$ for different values of the incident wave energy, $E=0$, 0.5 and 1, in the absence of saturation effects ($\beta=0$). The value of the nonlinearity parameter $\alpha$ is equal to 0.5 and the disorder strength $W$ is equal to 2. For small values of the incident wave intensity, the localization length is a decreasing function of it. For a sufficiently large $\vert r_0\vert^2$, however, the localization length is found to approach a saturation value.}
\end{figure}

In order to calculate the localization length $\xi$, we need to obtain the disorder-averaged logarithm of the transmittance $\langle \ln T \rangle$.  We have computed  $\langle \ln T \rangle$ by averaging over 10,000 random configurations of $\epsilon_n$. Calculations were performed for the system size $L$ up to 60. In the linear case, it is necessary to do the calculation for substantially larger values of $L$ in order to extract accurate results for the localization length. In the nonlinear case, however, the exponential decay of $\langle T \rangle$ or the linear decay of $\langle \ln T \rangle$ with $L$ is achieved for much smaller values of $L$. In the present
study, we have verified numerically that calculations up to $L=60$ are sufficient and $\langle \ln T \rangle$ approaches values smaller than $-6$ at this value of $L$. Our main aim is to investigate the effects of nonlinearity on localization properties and all of our results were obtained for a fixed disorder strength $W=2$. The step size for $t$ was $\delta=10^{-7}$. The error in the calculated value of $\vert r_0\vert^2$ was always smaller than $10^{-5}$.

\begin{figure}
\resizebox{0.7\textwidth}{!}{%
\includegraphics{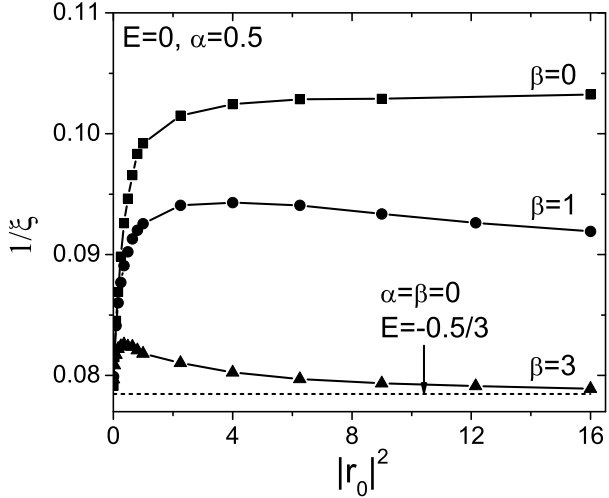}}
\caption{Inverse localization length $1/\xi$ plotted as a function of the incident wave intensity, $\vert r_0\vert^2$, when $E=0$, $\alpha=0.5$ and $\beta=0$, 1, 3. When $\vert r_0\vert^2$ or $\beta$ is sufficiently large, an effective linear behavior is achieved. The curve for $\beta=3$ is compared with the dashed straight line corresponding to the linear result obtained when $E=-0.5/3$.}
\end{figure}

In Fig.~2, we plot the inverse localization length $1/\xi$ as a function of the incident wave intensity $\vert r_0\vert^2$ in the absence of saturation effects ($\beta=0$). The energy of the incident wave, $E$, is equal to 0, 0.5 and 1 and the nonlinearity parameter $\alpha$ is fixed to 0.5. We find that the enhancement of localization occurs in the presence of weak nonlinearity for all cases. When $\vert r_0\vert^2$ is small, the localization length is a decreasing function of it. For sufficiently large values of $\vert r_0\vert^2$, the localization length is found to approach a saturation value. The specific saturation value of $\xi$ depends on the energy $E$ and the disorder strength $W$. A similar behavior has also been demonstrated in our previous work, though there is a subtle difference in the way in which the results are obtained \cite{Ng1}. In contrast to the present study, in \cite{Ng1}, the incident wave intensity $\vert r_0\vert^2$ was fixed and the localization length was calculated as a function of the nonlinearity parameter $\alpha$. When $\beta$ is zero, varying $\vert r_0\vert^2$ is theoretically equivalent to varying $\alpha$. When $\beta$ is nonzero, however, they are not equivalent.

\begin{figure}
\resizebox{0.7\textwidth}{!}{%
\includegraphics{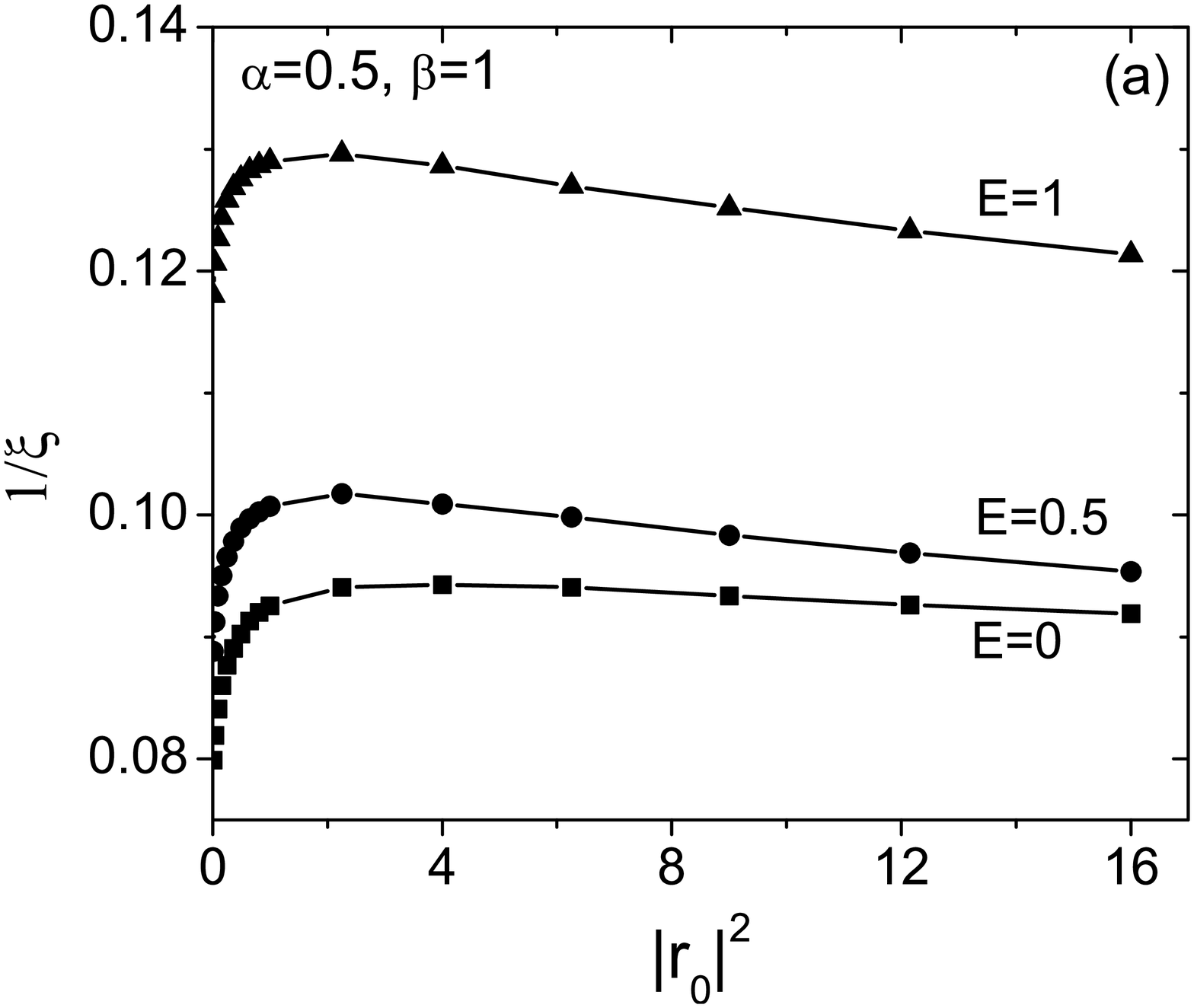}}
\resizebox{0.7\textwidth}{!}{%
\includegraphics{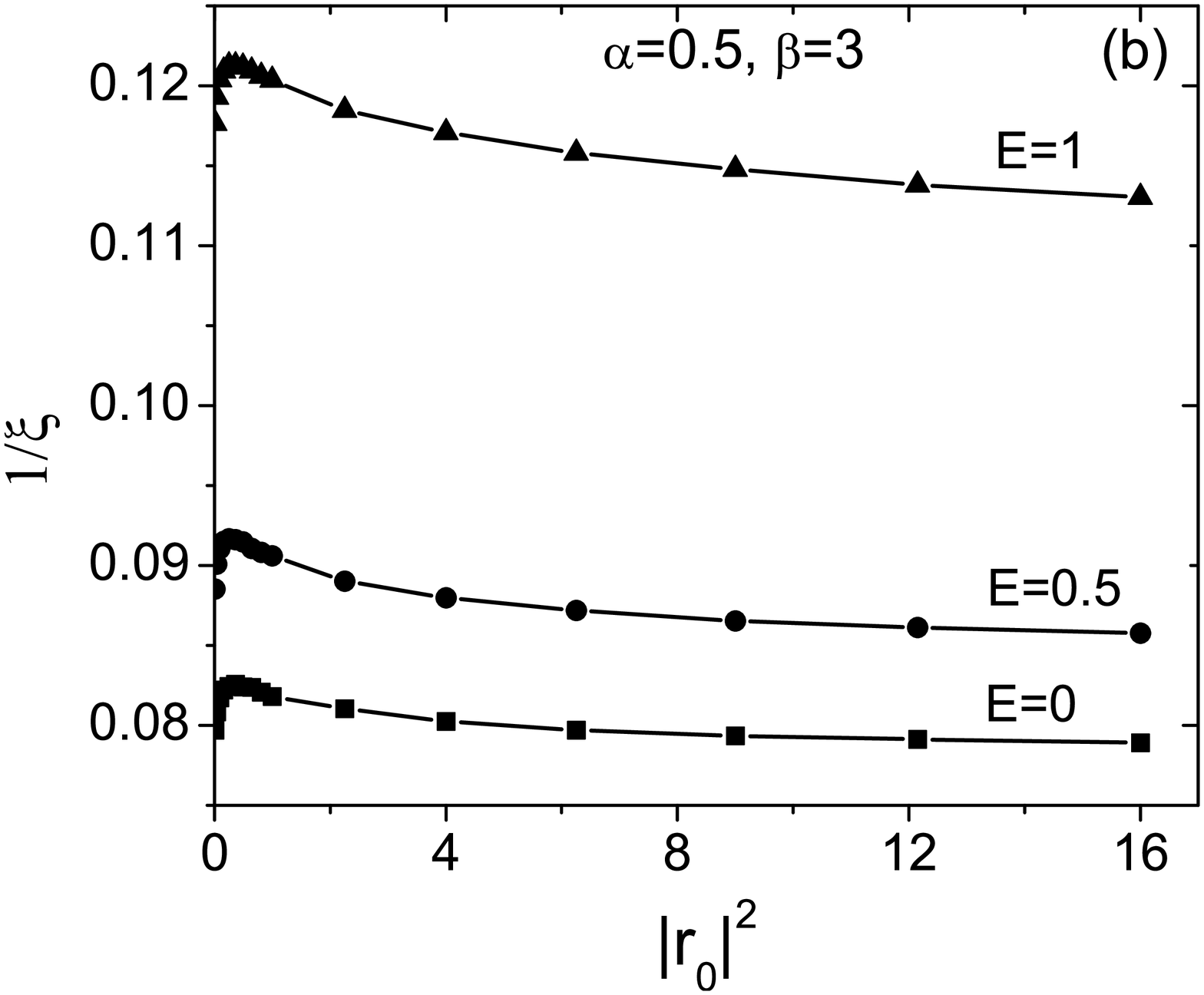}}
\caption{Inverse localization length $1/\xi$ plotted as a function of the incident wave intensity $\vert r_0\vert^2$ for three values of the energy, $E=0$, 0.5, 1, and $\alpha=0.5$, when (a) $\beta=1$ and (b) $\beta=3$. The nonlinear saturation effect depends strongly on the energy of the incident wave and is more pronounced when $E$ is large.}
\end{figure}

\begin{figure}
\resizebox{0.7\textwidth}{!}{%
\includegraphics{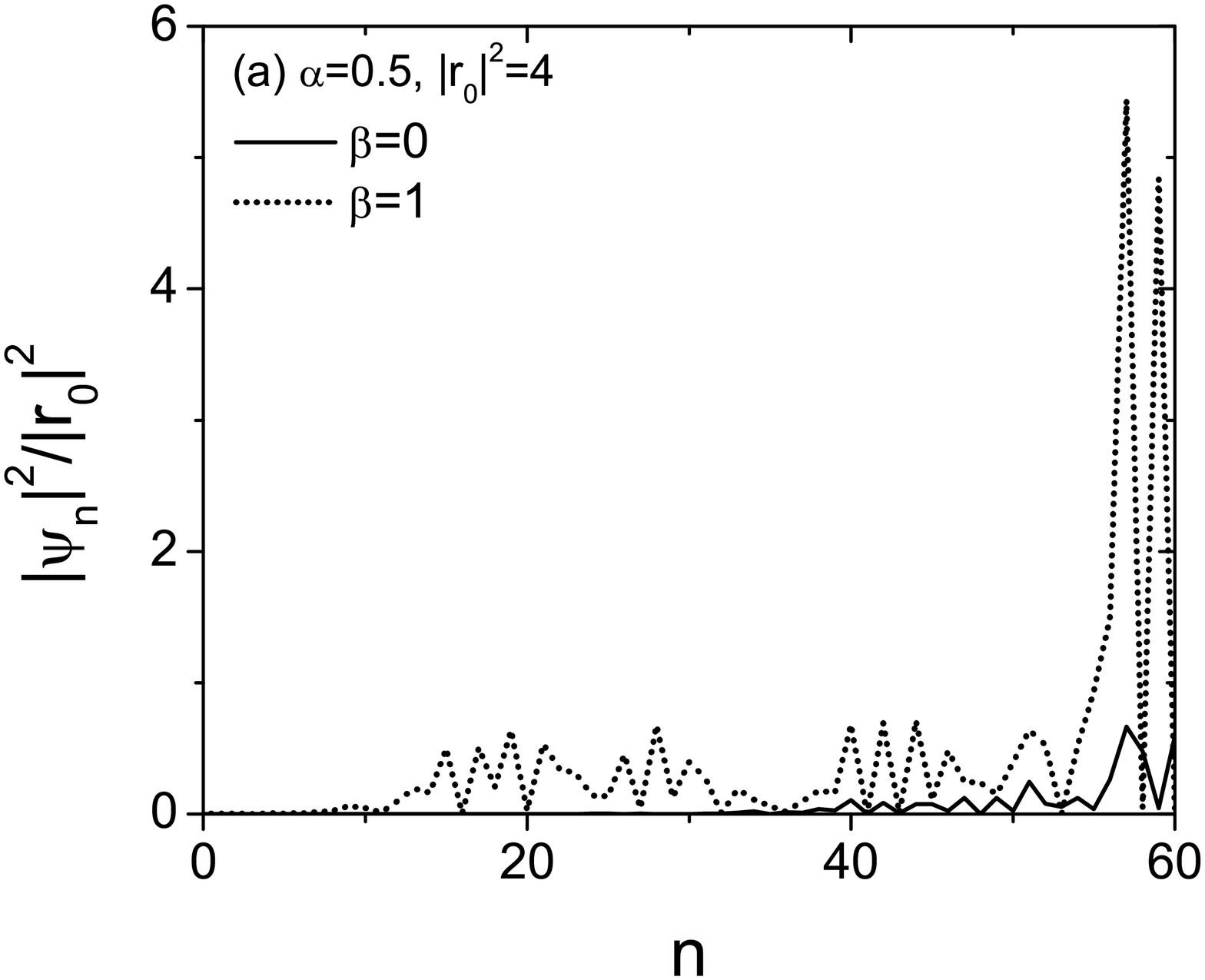}}
\resizebox{0.7\textwidth}{!}{%
\includegraphics{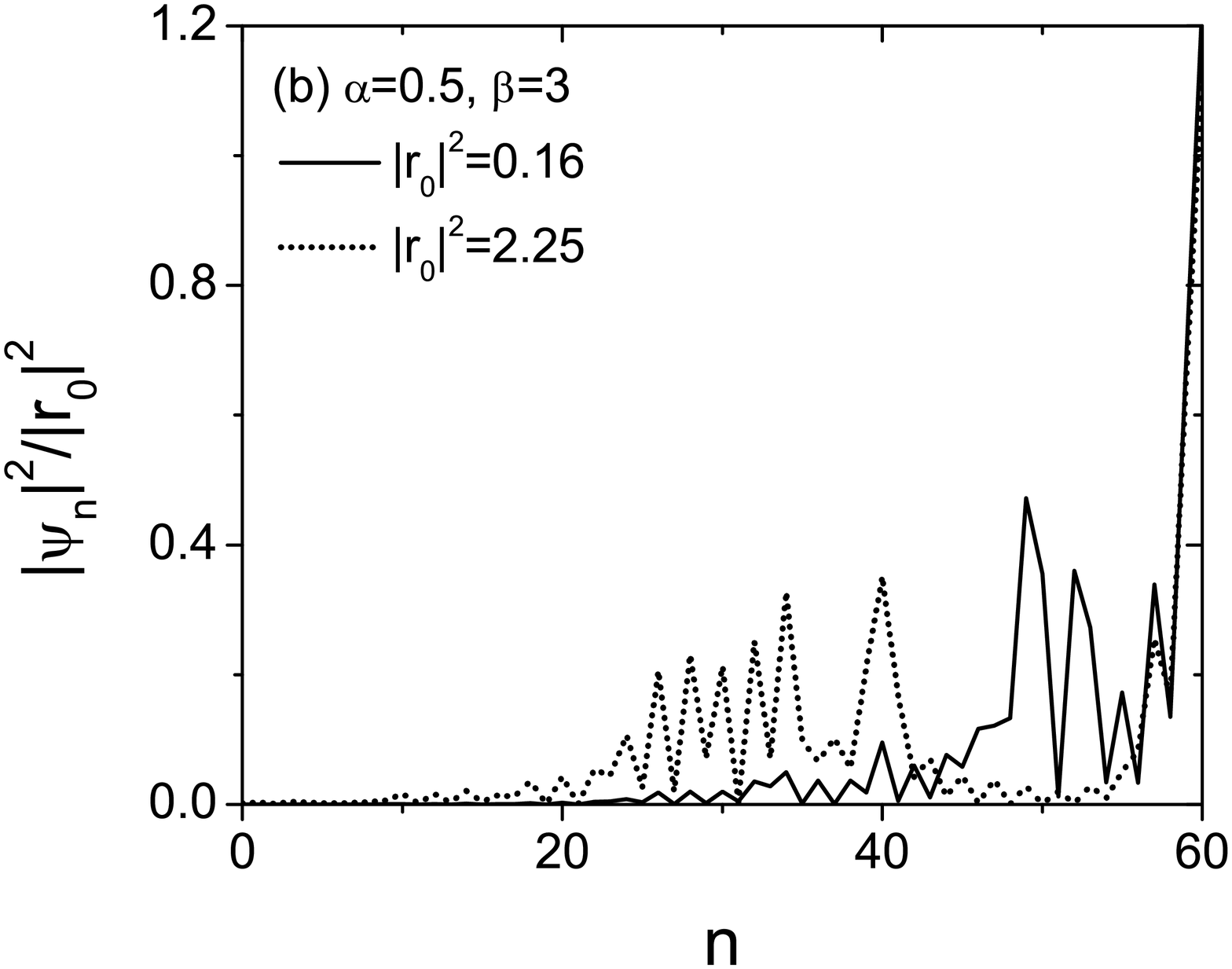}}
\caption{Spatial distribution of the intensity of the normalized wave function for one particular realization of the random potential, when (a) $\alpha=0.5$, $\beta=0$, 1, $\vert r_0\vert^2=4$ and (b) $\alpha=0.5$, $\beta=3$, $\vert r_0\vert^2=0.16$, 2.25. The values of $N$ and $E$ are fixed to 60 and 0 respectively. A plane wave is assumed to be incident from the right side.}
\end{figure}

In Fig.~3, we plot the inverse localization length $1/\xi$ as a function of the incident wave intensity $\vert r_0\vert^2$, when $E=0$, $\alpha=0.5$ and $\beta=0$, 1, 3. When the saturation of the nonlinear response is taken into account ($\beta \neq 0$), the localization behavior of waves is substantially changed. A nonmonotonic dependence of the localization length on the incident wave intensity occurs for finite saturation strengths. The inverse localization length takes a maximum value, which is a decreasing function of $\beta$, at a certain value of $\vert r_0\vert^2$. For low incident intensities where the saturation effect is ineffective, the enhancement of localization due to nonlinearity still occurs in a way similar to the case with $\beta=0$. For sufficiently high incident intensities, however, the inverse localization length is a decreasing function of the incident wave intensity, which implies that localization is suppressed for stronger input intensities, and ultimately approaches a saturation value. This feature is associated with the fact that the nonlinear system is reduced to an effectively linear one with
$\alpha=\beta=0$ and the renormalized energy $\tilde{E}=E-\alpha/\beta$ due to the saturation of the nonlinear response, when either $\vert r_0\vert^2$ or $\beta$ is sufficiently large. We illustrate this by comparing the curve for $\beta=3$ with the dashed straight line corresponding to the linear result obtained when $E=-0.5/3$ in Fig.~3.

In a recent paper, where the wave propagation through a saturable nonlinear asymmetric dimer has
been studied, it has been shown that the saturation of the nonlinear response has distinct impacts on the wave transmission properties for short-and long-wavelength input signals \cite{Ass}. In other words, the saturation effect depends strongly on the energy of incident waves. A similar behavior is also observed in the present study. In Fig.~4, we plot the inverse localization length as a function of the incident wave intensity for three values of the energy, $E=0$, 0.5, 1, and $\alpha=0.5$, when $\beta$ is equal to 1 and 3. From both figures, we observe that the nonlinear saturation effect is stronger and more pronounced when the energy of the incident wave $E$ is larger.


In Fig.~5, we show some examples of the spatial distribution of the intensity of the wave function calculated for a particular realization of disorder, when the system size $N$ is 60 and a plane wave is incident from the region where $n>60$. The wave function intensity is normalized by the intensity of the incident wave, $\vert r_0\vert^2$. The parameters $\alpha$ and $E$ are fixed to 0.5 and 0 respectively. In Fig.~5(a), we fix
$\vert r_0\vert^2=4$ and compare the results for $\beta=0$ and $\beta=1$. We find that the wave penetrates more deeply into the medium and its intensity is significantly larger in a wider region when $\beta=1$. This is consistent with the result that the localization length for $\beta=1$ is
larger than that for $\beta=0$ (see Fig.~3). In Fig.~5(b), we fix
$\beta=3$ and compare the results for $\vert r_0\vert^2=0.16$ and $\vert r_0\vert^2=2.25$. Again, we find that the wave penetrates more deeply into the medium when $\vert r_0\vert^2=2.25$, for which the localization length is
larger as can be seen in Fig.~4(b).

The effects studied in this work may be realized in experiments on layered photonic structures. In such systems, it has been demonstrated that the evolution of longitudinal Bloch waves can be approximated sensibly by the discrete nonlinear Schr\"{o}dinger equation \cite{Hen,Kos}.
For instance, we can employ a layered photonic system that is fabricated from lithium niobate ($\rm LiNbO_3$) crystals. This type of material exhibits a saturable self-defocusing nonlinearity corresponding to our consideration, via photo-refractive effects \cite{Kip,Che}. Spatial disorder is introduced into the system through a random variation of the layer width in the process of fabrication.

\section{Conclusion}

In this work, we have presented a numerical study of the propagation and the Anderson localization of plane waves in a 1D lattice chain, where  disorder and saturable nonlinearity are simultaneously present. Using a calculation scheme for solving a disordered version of the stationary discrete nonlinear Schr\"{o}dinger equation in the fixed input case, the disorder-averaged logarithmic transmittance and the localization length have been calculated in a numerically precise manner. For low incident intensities where the saturation effect is ineffective, the enhancement of localization due to nonlinearity has been found to occur in a way similar to the case without saturation. For sufficiently high incident intensities, we have found that the localization length is an increasing function of the incident wave intensity, which implies that disorder-induced localization is suppressed for stronger input intensities. This feature is associated with the fact that the nonlinear system is reduced to an effectively linear one, when either the incident wave intensity or the saturation parameter is sufficiently large.
We have also found that the nonlinear saturation effect is stronger and more pronounced when the energy of the incident wave is larger.





\section*{Acknowledgments}

This research is funded by Vietnam National Foundation for Science and Technology Development (NAFOSTED) under grant number 103.01-2014.10.
It is also supported by a National Research Foundation of Korea Grant (NRF-2015R1A2A2A01003494) funded by the Korean Government.

\end{document}